\documentclass[a4paper,amsmath,amssymb]{jpconf}
\usepackage{graphicx}
\usepackage{epstopdf}
\usepackage{amsmath, amsthm, amssymb}
\begin{document}
\title{Rheology and dynamical heterogeneity in frictionless beads at jamming density}
\author{Takahiro Hatano}
\address{Earthquake Research Institute, University of Tokyo, Tokyo 113-0032, Japan}
\ead{hatano@eri.u-tokyo.ac.jp}

\begin{abstract}
We investigate the rheological properties of an assembly of inelastic (but frictionless) particles close to the jamming density using numerical simulation, in which uniform steady states with a constant shear rate $\dot\gamma$ is realized. The system behaves as a power-law fluid and the relevant exponents are estimated; e.g., the shear stress is proportional to $\dot\gamma^{1/\delta_S}$, where $1/\delta_S=0.64(2)$. It is also found that the relaxation time $\tau$ and the correlation length $\xi$ of the velocity increase obeying power laws: $\tau\sim\dot\gamma^{-\beta}$ and $\xi\sim\dot\gamma^{-\alpha}$, where $\beta=0.27(3)$ and $\alpha=0.23(3)$.
\end{abstract}

\section{Introduction}
\label{sec:Introduction}
The rheological properties of suspensions are important in various contexts ranging from industries to earth science. A certain class of suspensions may be modeled as repulsive beads in a medium. For systems of a very low Reynolds number, long-range hydrodynamic interactions may be ignored and thus model suspensions may be further simplified as an assembly of repulsive particles at zero temperature and no gravity. Such repulsive sphere models are also used as models for granular matter or a certain class of foams \cite{durian}. In such simple models, the important parameter that determines the rheology is the density. In particular, it is well known that the viscosity seems to diverge at a certain density. Furthermore, the repulsive particles are inevitably in contact above a certain density and the system acquires rigidity without crystallization \cite{aharonov,ohern1,ohern2}, which is now referred to as the jamming transition.

It has been observed in many systems that the length and the time scales are observed to increase as the density approaches the critical jamming density \cite{drocco,dauchot,durian1,durian2,durian3,lechenault,olsson,hatano3}.
Importantly, however, such long wavelength fluctuations are observable only in a dynamic quantity 
such as the displacement of grains during a certain time lag \cite{lechenault}.
This spatial heterogeneity is due to the cooperative nature of the particle rearrangement. Quite interestingly, a similar heterogeneity is also observed in supercooled liquids \cite{kob,yamamoto}, 
in which the correlation length and time increases as the temperature decreases or the density increases. Although the relation between these heterogeneities in very different systems is not clear, they may lead to an overarching concept in general amorphous systems.

It is recognized that the cooperative motion of particles also plays an essential role in the rheological properties \cite{jenkins,mitarai,bizon}. However, to this date, we still do not understand the nature of the cooperative motion and how it affects dense granular rheology.
In this paper, by molecular dynamics simulation, we investigate the nature of velocity correlation 
in dense granular flow at the jamming density. In particular, we define the correlation length and time to find that they grow as the shear rate decreases. We also show that they dominate the dynamic properties of dense granular flow, particularly diffusion and rheology.

\section{Model}
Here we consider a monodisperse particulate system, where the particle diamater and the mass is denoted by $d$ and $M$, respectively. The volume fraction is fixed to be $0.639$, which is the jamming density estimated for three dimensional systems within the precision of $\pm0.001$ \cite{ohern2}. The dimensions of the system is $L\times L\times L$, where we set $L=40.32$ ($80000$ particles) unless otherwise indicated. We set the flow direction along the $x$ axis and the velocity gradient along the $y$ axis, and adopt Lees-Edwards boundary conditions \cite{allen}.

The radius and the position of particle $i$ are denoted by $R_i$ and ${\bf r}_{i}$, respectively.
Then using ${\bf r}_{ij}:={\bf r}_{i}-{\bf r}_{j}$, ${\bf n}_{ij}:={\bf r}_{ij}/|{\bf r}_{ij}|$, and $h_{ij} := (R_i+R_j) -|{\bf r}_{ij}|$, the repulsive force acting between particles $i$ and $j$ is written as ${\bf F}_{ij} := \left[kh_{ij} - \zeta{\bf n}_{ij}\cdot\dot{{\bf r}}_{ij}\right]{\bf n}_{ij}$.
If $R_i+R_j < |{\bf r}_{ij}|$, particles $i$ and $j$ are not in contact so that the force vanishes.
The coefficient of restitution, $e$, is given via the relation
\begin{equation}
e=\exp\left[-\frac{\pi}{\sqrt{4\bar{M}k/\zeta-1}}\right],
\end{equation}
where $\bar{M}$ is the reduced mass so that $\bar{M}=M/2$ in a monodisperse system.
We choose $2Mk/\zeta = 1$, which corresponds to the vanishing coefficient of restitution.
(We intended to model soft and viscous particles like tapioca).
Throughout this study, we adopt the units in which $d=1$, $M=1$, and $k=1$. (thus $\zeta=2$.)

A constant shear rate $\dot\gamma$ is applied to the system through the Lees-Edwards boundary conditions \cite{allen}.
Note that under these boundary conditions the system volume is constant.
Thus, the important parameters here are the shear rate $\dot\gamma$ and the packing fraction denoted by $\phi$.
A steady state of uniform shear rate $\dot\gamma$ can be realized starting from a class of special initial conditions. Here we investigate such uniform steady states.
Accordingly, it is convenient to define in the following discussions the velocity fluctuation with respect to steady shear flow. The velocity fluctuation is defined by ${\bf v}_i := {\bf r}_i - \dot\gamma {\bf n}_y \cdot {\bf r}_i$, where ${\bf n}_y$ is the unit vector along the $y$ axis.

\section{Velocity autocorrelation}
\begin{figure}
\begin{center}
\includegraphics[width=14cm]{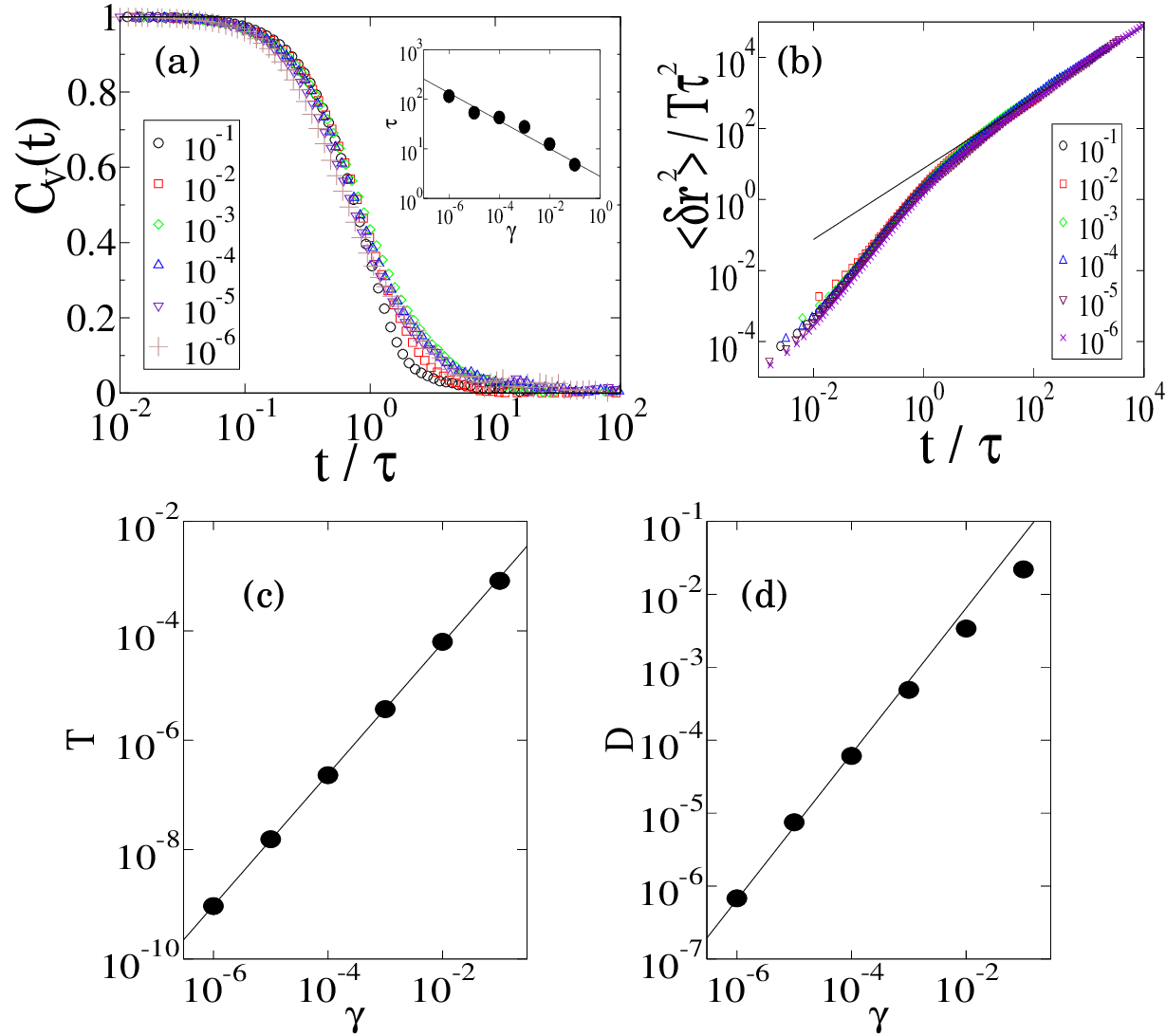}
\end{center}
\caption{\label{tauV} \footnotesize 
(a) Collapse of the velocity autocorrelation function $C_V(t)$ at various shear rates 
by rescaling the time with $\tau$. The legends show the shear rate.
The inset shows $\tau$ as a function of the shear rate,
where the solid line is proportional to $\dot\gamma^{-0.27}$.
(b) Collapse of the diffusion profiles at various shear rate.
Here the mean square displacement $\langle \delta {\bf r}_i^2(t)\rangle $ is normalized 
by the kinetic temperature $T\equiv \langle {\bf v}^2_i \rangle$ and $\tau^2$.
This means that the diffusion coefficient is scaled with $T\tau$.
The solid line is proportional to $t$.
(c) The shear-rate dependence of the kinetic temperature.
The solid line is proportional to $\dot\gamma^{1.2}$.
(d) The shear-rate dependence of the diffusion coefficient.
The solid line is proportional to $\dot\gamma^{1.0}$.}
\end{figure}

First we investigate the cooperative nature of the velocity fluctuation ${\bf v}_i(t)$ by defining the velocity autocorrelation function of a tracer particle $i$.
\begin{equation}
\label{vvdefinition}
C_V(t) = \frac{\langle {\bf v}_i(t)\cdot{\bf v}_i(0)\rangle}{\langle{\bf v}_i^2\rangle}.
\end{equation}
This correlation function can be calculated at several shear rates ranging  $\dot\gamma=10^{-1}$ to $10^{-6}$.
At each shear rate the correlation function exhibits a characteristic relaxation time denoted by $\tau$. We estimate this time scale by rescaling the time with $\tau(\dot\gamma)$ so as to $C_V(t/\tau)$ at each shear rate collapses. Such collapse is shown in Figure \ref{tauV} (a).
Recalling that the system is marginally jammed (i.e., the mean free path is zero), it may be remarked that the velocity correlation time is several times larger than $\sqrt{k/M}=1$. Furthermore, this characteristic time increases obeying a power law as the shear rate decreases;
\begin{equation}
\label{timescale}
\tau\propto\dot\gamma^{-\beta},
\end{equation}
where $\beta=0.27(3)$. See the inset of Figure \ref{tauV} (a).
It is also noteworthy that the relaxation time is still much smaller than $\dot\gamma^{-1}$, which is the structural relaxation time due to shear flow. We can thus regard $\tau$ as the intermediate time scale at sufficiently low shear rates; i.e., $\sqrt{m/k} \ll \tau \ll \dot\gamma^{-1}$.

The velocity correlation time involves the nature of diffusion.
The diffusion profile is defined by $\langle\delta{\bf r}_i^2(t)\rangle=|\int_0^t ds {\bf v}_i(s)|^2$. We compute this quantity at several shear rates and find that they collapse using the velocity correlation time $\tau$ and the kinetic temperature defined by $T:=\langle {\bf v}^2_i \rangle$. Namely,
\begin{equation}
\label{diffusionscaling}
\langle\delta{\bf r}_i^2(\tau)\rangle = T\tau^2 f(t/\tau),
\end{equation}
where $f(\cdot)$ is a universal function. The scaling collapse is shown in the inset of Figure \ref{tauV} (b). One can immediately conclude from this scaling that the diffusion coefficient $D$ is scaled with $T\tau$.
As shown in Figure \ref{tauV} (c), the shear-rate dependence of the kinetic temperature is expressed as $T\propto\dot\gamma^{y_T}$, where $T=1.20(3)$.
Thus, $D\sim T\tau$ is expected to be proportional to the shear rate.
We estimate $D$ from the diffusion profile at each shear rate and find that $D$ is proportional to $\dot\gamma$ at sufficiently low shear rates ($\dot\gamma\le 10^{-3}$).
This is shown in Figure \ref{tauV} (d), where the solid line is proportional to $\dot\gamma$.
Note that the scaling of the diffusion coefficient is apparently anomalous at higher shear rates; it appears that $D\propto\dot\gamma^{y_D}$, where $y_D\simeq0.8$.
We remark that this anomalous scaling with almost the same exponent at higher shear rates has been observed in some similar systems such as a model foam under shear \cite{olsson3} and an experiment on colloidal system \cite{besseling}.

One may notice in Figure \ref{tauV} (b) that diffusion occurs after sufficiently long time (i.e., $f(x)\propto x$ for $x \gg1$), whereas superdiffusion occurs in the shorter time scale ($t\le\tau$).
It is important to notice that subdiffusion is not observed in contrast to other granular systems without flow \cite{durian1,durian2,durian3,lechenault}.
This indicates that the in-cage motion is not relevant to the present system and the structural relation takes place even in a shorter time scale $t\sim\tau$.
This is because the applied shear breaks the cages.
This enables us to detect the dynamical heterogeneity using a two-point correlation function instead of a four-point correlation function that is typically utilized to observe dynamical heterogeneities. This is discussed in the next section.

Another important quantity regarding the diffusion profile is the distance over which grains move cooperatively. This length scale is defined by $\sqrt{\langle\delta{\bf r}_i^2(\tau)\rangle}$ and denoted by $l_c$.
As is apparent from Eq. (\ref{diffusionscaling}), $l_c$ is scaled with $\sqrt{T}\tau$.
We confirm that this length scale is scaled with $\dot\gamma^{x}$ ($x\simeq0.3$) and thus vanishes in the zero shear rate limit.
This makes a quite contrast to supercooled liquids, in which $l_c$ is almost constant and as large as the cage size \cite{kob2}.
Because $l_c$ vanishes in the $\dot\gamma\rightarrow0$ limit, it is expected that superdiffusion does not occur (i.e., diffusion is observed at all time scales) at sufficiently slow shear rates \cite{heussinger}.


\section{Spatial heterogeneity of velocity}
\begin{figure}
\begin{center}
\includegraphics[scale=0.6]{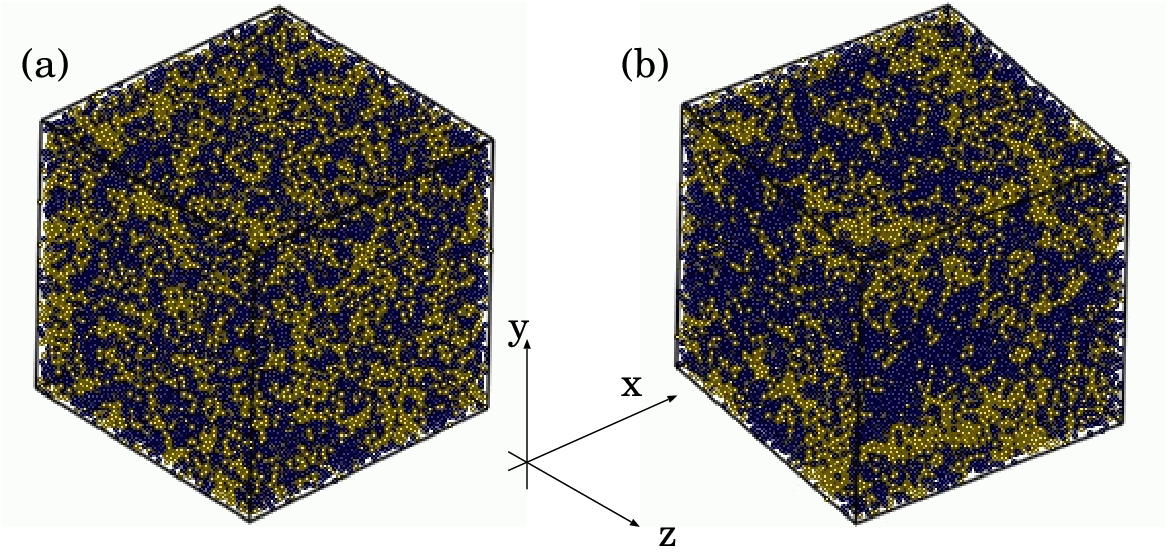}
\caption{\label{DHs}
Heterogeneity of the velocity fluctuation defined by equation (\ref{mi}).
Grains with larger velocity ($m_i < e^{-1}$) are brightly shown.
(a) A snapshot at $\dot\gamma=10^{-2}$.
(b) A snapshot at $\dot\gamma=10^{-5}$, where grains with large velocity form a lager cluster.}
\end{center}
\end{figure}

\begin{figure}
\begin{center}
\includegraphics[scale=0.6]{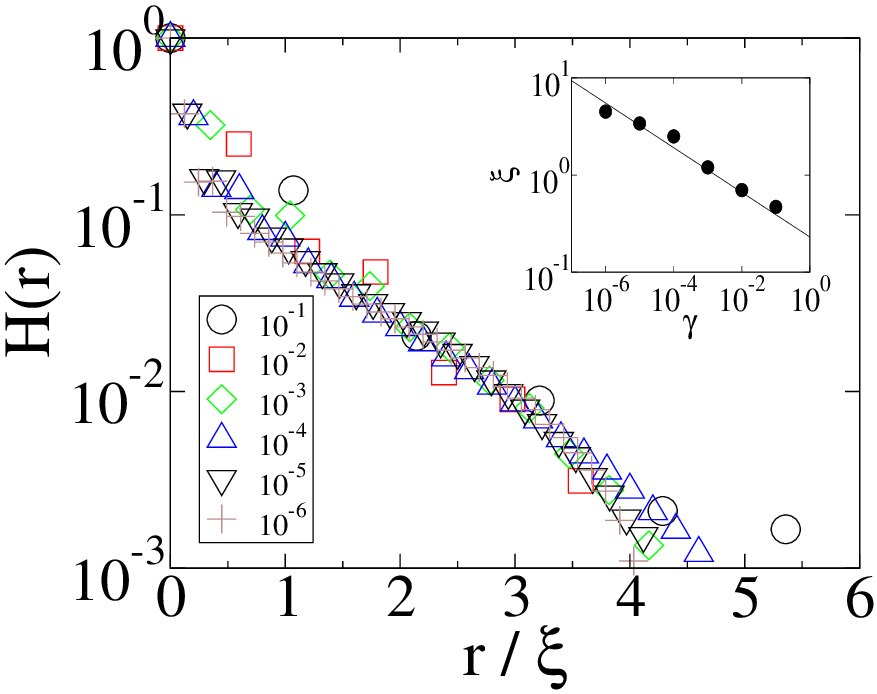}
\end{center}
\caption{\label{Hscaled}
The correlation functions defined by Eqs. (\ref{mi}) and (\ref{h}).
The legends indicate the shear rate.
Note that the length scale is normalized by $\xi$ so that the correlation functions collapse.
(Inset) The correlation length $\xi$ as a function of $\dot\gamma$. The solid line is proportional to $\dot\gamma^{-0.23}$.}
\end{figure}

The fluctuating velocity field ${\bf v}_i(t)$ exhibits spatial heterogeneity, which is enhanced at the slower shear rates.
To quantify the spatial heterogeneity of ${\bf v}_i(t)$ clearly, it is indeed convenient to define the following quantity rather than the velocity itself.
\begin{equation}
\label{mi}
m_i(t) \equiv \exp\left[-\frac{{\bf p}_i^2(t)}{\langle{\bf p}_i^2(t)\rangle_i}\right],
\end{equation}
where $\langle \cdot\rangle_i$ denotes the average over all grains. 
The snapshot of $m_i(t)$ is shown in Figure \ref{DHs}, where the growing correlation length 
at lower shear rate is apparent. 
Then we define the correlation function as 
\begin{equation}
\label{h}
h(r; t) \equiv \frac{\sum_{i\ge j} \delta m_i(t) \delta m_j(t)
\delta(r-|{\bf x}_i(t)-{\bf x}_j(t)|)}
{\sum_{i\ge j} \delta(r-|{\bf x}_i(t)-{\bf x}_j(t)|)}, 
\end{equation}
where $\delta m_i(t)\equiv m_i(t)-\langle m_i(t) \rangle_i$.
We remark that $h(r; t)$ does not significantly fluctuate in time in the present system 
(containing $80000$ grains) so that we drop the argument $t$ in $h(r; t)$ to write it as $H(r)$.
The correlation function $H(r)$ at various shear rates is shown in Figure \ref{Hscaled}.
This clearly shows that the correlation length $\xi$ increases as 
\begin{equation}
\label{length}
\xi \propto \dot\gamma^{-\alpha}, 
\end{equation}
where the exponent is estimated as $\alpha=0.23(3)$.
The power-law divergence of the correlation length with respect to the shear rate is consistent with the fact that the critical point is located at $\dot\gamma=0$ \cite{liu,ohern1}.
Note that equations (\ref{mi}) and (\ref{h}) respresent a two-point correlation.
This makes a quite contrast to vibrated \cite{lechenault} or air-fluidized granular media 
\cite{durian1,durian2,durian3} and other thermal glass formers such as supercooled liquids
\cite{yamamoto}, where the spatial heterogeneity can be detected only in a four-point correlation.
As discussed in the previous section, this may be one of the essential features of an athermal system under flow, in which the cage is washed away by the flow and therefore the kinetic arrest is absent.

Taking Eqs. (\ref{timescale}) and (\ref{length}) into account, the time constant of the cooperative motion is scaled with $\tau\propto\xi^z$, where $z=1.3(1)$.
Interestingly, this is comparable to an experiment on a two dimensional air-fluidized system \cite{durian3}.

We remark that the correlation length that diverges as $\xi\propto S^{-0.5}$ is also found in a simulation on a two-dimensional model foam \cite{olsson}. This indeed leads to $\xi\sim\dot\gamma^{-0.21}$ as they obtain $S\propto\dot\gamma^{0.42}$ at the critical density.
The exponent for the correlation length is coincident with that in the present study, although the models and the definition of correlation length are slightly different. This coincidence indicate that the exponent for the correlation length is insensitive to the dimensionality and the details of the model (i.e., viscosity of the medium and the nature of the dissipative force.)

We also remark that the correlation length is also found in an experiment on an inclined plane flow \cite{katsuragi}, in which the correlation length in terms of a four-point correlation increases as $V^{-0.25}$.
This also supports Eq. (\ref{length}) because the velocity profile is exponential so that $\dot\gamma\propto V$.
The important implication of this experiment is the insensitivity of the exponent for the growing correlation length $0.25$ on the nature of the interparticle force; i.e., whether it is frictional or frictionless.

\section{Rheology}
\subsection{Granular matter as a power-law fluid}
\begin{figure}
\begin{center}
\includegraphics[scale=0.8]{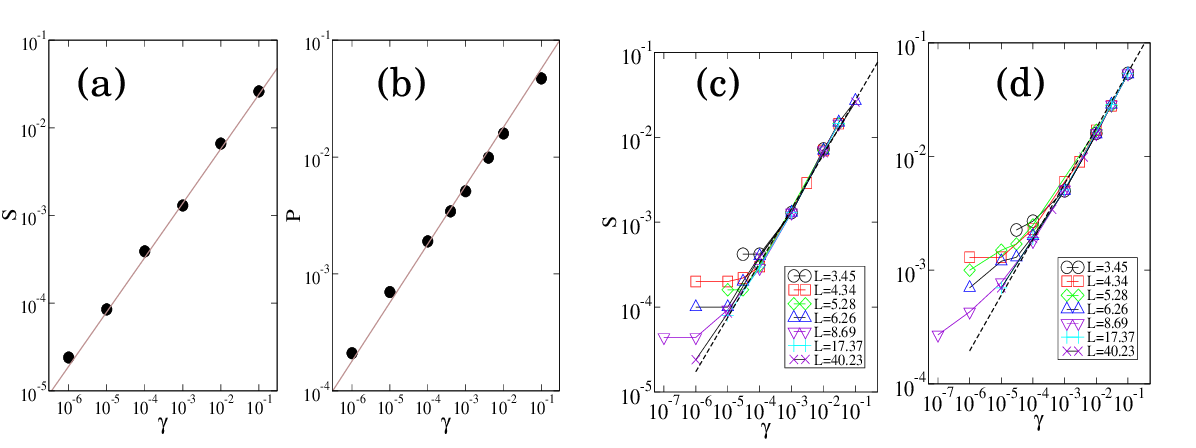}
\caption{\label{rheology}
(a) The shear-rate dependence of the shear stress.
The solid line is proportional to $\dot\gamma^{0.64}$.
(b) The shear-rate dependence of the pressure.
The solid line is proportional to $\dot\gamma^{0.5}$.
(c) System-size dependence of the shear stress.
The dashed line is proportional to $\dot\gamma^{0.64}$.
(d) System-size dependence of the pressure.
The dashed line is proportional to $\dot\gamma^{0.5}$.}
\end{center}
\end{figure}

Next we discuss the rheology. As shown in Figure \ref{rheology}, the shear stress and the pressure exhibit power-law dependences on the shear rate: $S\propto\dot\gamma^{1/\delta_S}$ and $P\propto\dot\gamma^{1/\delta_P}$, where $1/\delta_S=0.64\pm0.02$ and $1/\delta_P=0.5\pm0.02$.
Note that these exponents do not contradict those previously conjectured for a granular matter at the critical density \cite{hatano1,hatano2}.
These exponents differ from $2.0$, i.e., Bagnold's scaling \cite{bagnold}, which holds only in unjammed regime where the stiffness of the particles does not affect the mean free time. The nontrivial power-law rheology arises where the mean free time (or the mean free path) vanishes.
We remark that similar power-law behaviors of the shear stress have been observed in other amorphous particulate systems such as foam models \cite{olsson,tighe,olsson2}, although the observed exponent is different from the present value. In foam models, the exponent $1/\delta_S$ has been estimated as $0.28$ \cite{olsson2} or $1/2$ \cite{tighe}, which are significantly lower than that in the present model.
It has been shown that the rheological exponents are sensitive to the details of the force models \cite{ohern2,hatano1} and the nature of the medium \cite{olsson,hatano1}. We wish to remark that the foam models \cite{olsson,tighe,olsson2} do not posses inertia and this may make the difference in the rheological exponents.
However, notwithstanding a plausible theory on the rheological exponents \cite{tighe}, there is yet no consensus regarding their precise values.

\subsection{Finite-size effect}
\begin{figure}
\begin{center}
\includegraphics[scale=0.4]{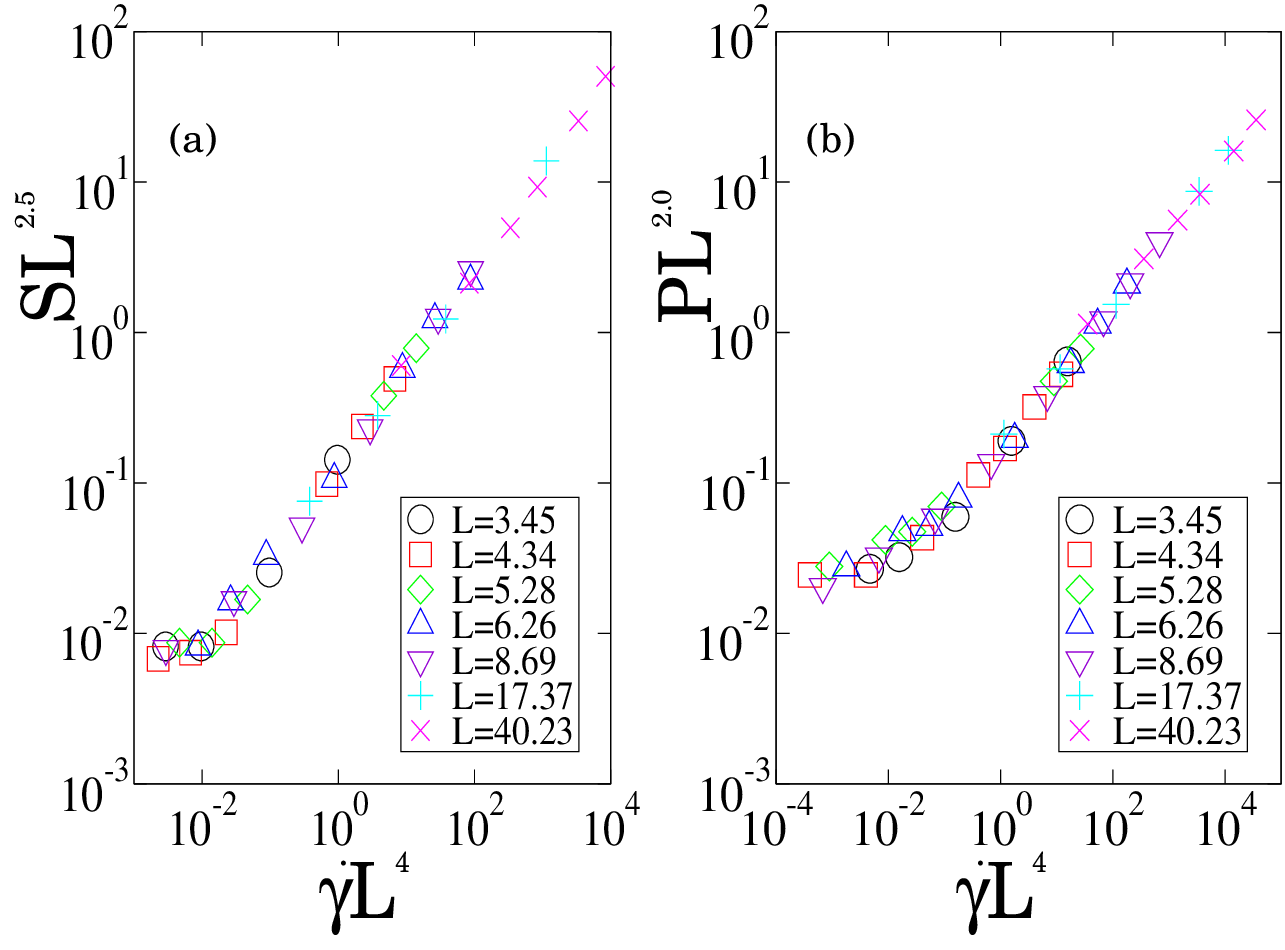}
\caption{\label{scaling} Collapse of the rheological data shown in Figure. \ref{rheology} 
by finite-size scaling: (a) equation (\ref{scalingS}), and (b) equation (\ref{scalingP}).}
\end{center}
\end{figure}

Generally, the true critical behaviors cannot be observed in a finite system, 
as the growing correlation length eventually exceeds the system size.
To detect such a finite size effect, we investigate the system size dependence of rheology 
in the following way.
The control parameters here are the shear rate $\dot\gamma$ and the system size $L$ 
so that the shear stress $S$ and the pressure $P$ are formally expressed as 
$S=S(\dot\gamma, L^{-1})$ and $P=P(\dot\gamma, L^{-1})$.
These relations obtained in our simulation are shown in Figure \ref{rheology}.
Note that, at lower shear rates, the shear stress and the pressure are larger 
than expected from the power law behaviors.
This tendency is remarkable for smaller systems; i.e., 
the crossover occurs at higher shear rate as the system becomes smaller.
This system-size dependent rheology is described by finite size scaling, 
which are of the same form as those in conventional critical phenomena \cite{notescaling}.
\begin{eqnarray}
\label{scalingS}
S(\dot\gamma, L) &=& L^{-1/y_S}f_S(\dot\gamma L^{1/y_{\dot\gamma}}),\\
\label{scalingP}
P(\dot\gamma, L) &=& L^{-1/y_P}f_P(\dot\gamma L^{1/y_{\dot\gamma}}), 
\end{eqnarray}
where $f_S$ and $f_P$ are scaling functions.
Figure \ref{scaling} shows the collapse of the rheological data using Eqs. (\ref{scalingS}) 
and (\ref{scalingP}), where the exponents are estimated as 
$1/y_S=2.5\pm0.2$, $1/y_P=2.0\pm0.2$, and $1/y_{\dot\gamma}=4.0\pm 0.5$.
We can also see that the finite size effect comes into play where 
$\dot\gamma L^{1/y_{\dot\gamma}}\le 0.1$; i.e., $L \le \dot\gamma^{-y_{\dot\gamma}}$.
This means that the correlation length increases as $\xi\sim\dot\gamma^{-y_{\dot\gamma}}$,
because the finite size effect is generally due to the correlation length 
that exceeds the system size.
Thus, $y_{\dot\gamma}$ must be identical to $\alpha$ in equation (\ref{length}), 
which is estimated using a spatial correlation function, equations (\ref{mi}) and (\ref{h}).
This is indeed valid within the numerical precision.
Therefore we can conclude that the finite-size effect in rheology is due to the fact that 
the correlation length of the dynamical heterogeneity exceeds the system size.


\begin{thebibliography}{99}
\bibitem{notescaling}
In general, finite-size scaling reads 
$S(\Phi, \dot\gamma, L) = L^{-1/y_S} f_S(\Phi L^{1/y_{\Phi}}, \dot\gamma L^{1/y_{\dot\gamma}})$. 
where $\Phi$ is the distance from the critical density.
Thus, $\Phi$ must be zero (or $\Phi L^{1/y_{\Phi}}\simeq 0$) if one wishes to estimate $y_{\dot\gamma}$.
\end{thebibliography}
\end{document}